\documentclass[10pt]{article}
\usepackage{palatino}
\usepackage{amsmath}
\usepackage[a4paper, total={6.6in, 10in}]{geometry}
\usepackage{graphicx}
\usepackage[breaklinks=true,colorlinks=true,linkcolor=blue,urlcolor=blue,citecolor=blue]{hyperref}
\usepackage{abstract}
\renewcommand{\abstractname}{}

\usepackage{multicol}
\usepackage{float}
\usepackage{braket}
\usepackage{csquotes}

\usepackage{titlesec}

\titleformat*{\section}{\large\bfseries}
\titleformat*{\subsection}{\normalsize\bfseries}

\usepackage[font=footnotesize]{caption}

\usepackage{subfig}
\usepackage{cite}
\begin{document}
	
	\title{Hybrid Photonic Loss Resilient Entanglement Swapping}
	\date{}
	
	\author{Ryan C. Parker$^{1}$, Jaewoo Joo$^{2,3}$, Mohsen Razavi$^{2}$, Timothy P. Spiller$^{1}$}
		\maketitle
	
	\vspace{-1cm}
\begin{center}
	$^{1}$\textit{York Centre for Quantum Technologies, Department of Physics, University of York, York, YO10 5DD, U.K.}

$^{2}$\textit{School of Electronic and Electrical Engineering, University of Leeds, Leeds, LS2 9JT, U.K.}

$^{3}$\textit{School of Computational Sciences, Korea Institute for Advanced Study, Seoul 02455, Korea}
\end{center}

	\begin{abstract}
		
		We propose a scheme of loss resilient entanglement swapping between two distant parties in lossy optical fibre. In this scheme, Alice and Bob each begin with a pair of entangled non-classical states; these "hybrid states" of light are entangled discrete variable (Fock state) and continuous variable (coherent state) pairs. The continuous variable halves of each of these pairs are sent through lossy optical fibre to a middle location, where these states are then mixed (using a 50:50 beam-splitter) and measured. The detection scheme we use is to measure one of these modes via vacuum detection, and to measure the other mode using homodyne detection. 
		
		In this work we show that the Bell state $\ket{\Phi^{+}}=(\ket{00}+\ket{11})/\sqrt{2}$ can theoretically be produced following this scheme with high fidelity and entanglement, even when allowing for a small amount of loss. It can be shown that there is an optimal amplitude value ($\alpha$) of the coherent state when allowing for such loss. We also investigate the realistic circumstance when the loss is not balanced in the propagating modes. We demonstrate that a small amount of loss mismatch does not destroy the overall entanglement, thus demonstrating the physical practicality of this protocol. \\[0.08in]
		
	\end{abstract}

	\begin{multicols}{2}

	\section{Introduction}\label{intro}
	
	Distributing entanglement over long distances is a key enabler for quantum communications to be realised on a worldwide scale. Entanglement is an invaluable resource in quantum key distribution  \cite{Bennett1984,Ekert1991,Braunstein2012}, quantum secret sharing \cite{Gottesman2000,Tittel2001} and quantum teleportation \cite{Bennett1993,Braunstein1998}. Entanglement swapping is performed by two distant parties (Alice and Bob), that each possess a pair of entangled states (modes \enquote{$AB$} and \enquote{$CD$} respectively). If they each send one of their systems ($B$ and $D$) to a central location, a suitable joint measurement entangles the remaining systems ($A$ and $C$) that Alice and Bob still possess, thus the name \enquote{entanglement swapping} \cite{Zukowski1993}. Entanglement swapping (ES) in this way is analagous to a quantum teleportation scheme, where modes $B$ and $D$ are \enquote{teleported} to modes $A$ and $C$ respectively as a result of the joint measurement of modes $B$ and $D$ \cite{Bouwmeester2000}. 
	
	Currently, ES protocols suffer from sending quantum signals through an optical fibre which introduces decoherence and photon loss \cite{Gisin2002}. Mitigating against this issue takes ES protocols closer to practical implementation, with increased potential for application in quantum repeater \cite{Jin2015,Boone2015,Jones2016,Zwerger2016} and quantum relay \cite{Riedmatten2004,Collins2005,Ngah2014,Khalique2014} schemes. Furthermore, ES is a perfectly viable method of potentially realising truly long distance quantum communications \cite{Aspelmeyer2003,Riedmatten2005} and has recently been demonstrated at a distance of 100 km using optical fibre and time-bin entangled photon-pairs \cite{Sun2017}, and also at telecom wavelengths with high efficiency \cite{Jin2015}.
	
	ES was initially proposed using discrete variable (DV) states \cite{Zukowski1993}, and was shown experimentally using polarised photons \cite{Pan1998,Pan2001} and vacuum-one-photon quantum states \cite{Sciarrino2002}. However, as a result of detector inefficiencies lowering success probability (a Bell-State measurement is bounded by 1/2 when using only linear optical elements \cite{Calsamiglia2001}), ES events occur rarely when using only DVs. Research then began on the use of continuous variables (CVs) for ES \cite{Ralph1999,Abdi2004,Ban2004}, and was first performed experimentally in 2004 \cite{Jia2004}. Photonic coherent states work well for ES based on CV states, as coherent states are typically more resilient to photon losses \cite{Joo2016}. 
	
	In this paper we investigate the use of entangled hybrid states for application in an ES protocol. These hybrid states of light are entangled discrete and continuous variable quantum states. Hybrid states of light are particularly effective for ES schemes, and have been used in experimental proofs using squeezed states as the CV part\cite{Takeda2015} and also coherent states \cite{Brask2010}. The DV part uses as basis states the vacuum and single photon Fock (number) states, and the CV part uses the basis states of nearly orthogonal coherent states.
	
	This paper is organised as follows. In Section \ref{Method} we introduce the ES protocol used in this work, as well as the detection methods used. In Section \ref{Unequal} we introduce unequal lossy modes, and parametrise a value for this \enquote{loss mismatch}. In Section \ref{RandD} we show that the subsequent entanglement shared by Alice and Bob is not severely damaged when allowing for unequal lossy modes, and show that high levels of fidelity and entanglement can be reached. Our conclusions are given in Section \ref{Conclusions}. 
	
	\section{Entanglement Swapping with Loss}\label{Method}
	
	\subsection{Building Block Entangled States}
	We here use a specific bipartite entangled state (which we refer to as a hybrid entangled state), which has a DV qubit in a spatial mode and a CV qubit in the other mode, as follows:
	\begin{align}\label{hybrid}
	\ket{\psi_{HE}}_{AB}=\frac{1}{\sqrt{2}}(\ket{0}_{A}\ket{\alpha}_{B}+\ket{1}_{A}\ket{-\alpha}_{B}),
	\end{align}
	\noindent where the subscript $A$ and $B$ can be replaced with $C$ and $D$ respectively to describe the other initial hybrid entangled state $\ket{\psi_{HE}}_{CD}$. The mode $B$ is assumed to be a photonic coherent state going through a photon-lossy channel, while the stationary mode $A$ can be represented by various physical systems. For example, a hybrid photonic state has been recently demonstrated using a vacuum and a single-photon state for mode $A$ in \cite{Jeong14} as well as using polarisation photons in \cite{Laurat14}.
		
	Instead of the vacuum and single-photon states, for solid-state stationary qubits, atomic ensembles and ions can be excellent candidates to create the state $\ket{\psi_{HE}}$. For example, a non-maximally entangled state can be created in the hybrid fashion 
	\begin{align}
	\ket{\phi_{HE}}_{Ap} \approx \sqrt{1-p_c} \ket{G}_{A} \ket{0}_{p}+ \sqrt{p_c} \ket{W}_{A} \ket{1}_{p},
	\end{align}
	\noindent where $p_c$ is the success probability of having a single photon in spatial mode $p$, and $\ket{G}$ and $\ket{W}$ are the hyperfine states of an atomic ensemble (or an ion) \cite{DLCZ,Moonjoo,Moonjoo1}. Then, we build the optical set-up so that the spatial mode $B$ is matched with one of two directions of pair-wise parametric down-conversion photons from a nonlinear crystal, with efficiency $\eta$, while $\ket{\alpha}_B$ is injected along the other direction of the pair of photons. 
	\begin{align}
	\ket{\Psi^{tot}}_{ABp}&\approx \sqrt{1-\eta}\sqrt{1-p_c} \ket{G}_{A} \ket{0}_{p} \ket{\alpha}_{B} \nonumber\\
	&+\sqrt{\eta} \sqrt{1-p_c} \ket{G}_{A} \ket{1}_{p} a^+_B \ket{\alpha}_{B} \nonumber \\
	&+ \sqrt{1-\eta} \sqrt{p_c} \ket{W}_{A} \ket{1}_{p} \ket{\alpha}_{B}\nonumber\\
	&+ \sqrt{\eta}  \sqrt{p_c} \ket{W}_{A} \ket{2}_{p} a^+_B \ket{\alpha}_{B}	
	\end{align}
	If we detect a single photon in mode $p$ and $p_c=\eta$, the final state is approximately equal to 
	$(\ket{G}_{A} \ket{\alpha\rq{}}_{B}+\ket{W}_{A} \ket{-\alpha\rq{}}_{B})/\sqrt{2}$ \cite{Jeong14}.

 \subsection{Lossy Modes}	
	
	We use a vacuum state in mode $\varepsilon_{B}$  and $\varepsilon_{D}$ ($\ket{0}_{\varepsilon_{B}}$ and $\ket{0}_{\varepsilon_{D}}$ respectively) as is standard for modelling loss using a beam-splitter, where the second input state is the propagating coherent state in mode $B$ or $D$ which is mixed with the vacuum state to imitate loss. The two lossy modes are mixed at a 50:50 beam-splitter ($BS^{1/2}$) and are then measured using a vacuum projection in mode $B$ and a homodyne measurement in mode $D$. 
	
\begin{figure}[H]
		\centering
		\scalebox{0.35}{\includegraphics[trim={1cm 2cm 0 3cm},clip]{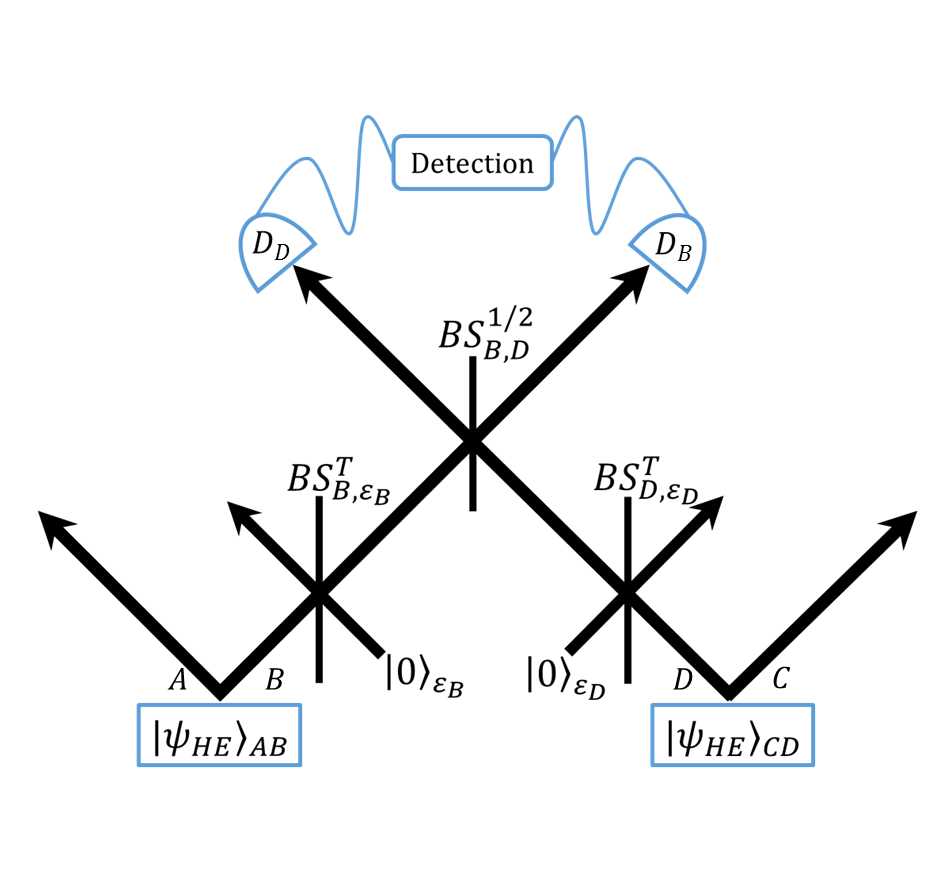}}
		\caption{Diagram to represent the four channel system (where $\ket{\psi_{HE}}_{AB}$ and $\ket{\psi_{HE}}_{CD}$ are entangled hybrid states) undergoing entanglement swapping with two lossy channels ($B$ and $D$), modelled by mixing a vacuum state ($\ket{0}_{\varepsilon_{B}}$ and $\ket{0}_{\varepsilon_{D}}$ respectively) using a beam-splitter of transmission rate $T$ ($BS^{T}_{B,\varepsilon_{B}}$ and $BS^{T}_{D,\varepsilon_{D}}$). The lossy modes $B$ and $D$ are then mixed at a 50:50 beam-splitter ($BS^{1/2}_{B,D}$) and subsequently measured ($D_{B}$ and $D_{D}$) to complete the protocol.}
		\label{fig:ES}
	\end{figure}

	The full ES protocol, including loss, is shown in Fig. \ref{fig:ES}.
	\noindent Through this ES protocol, Alice and Bob can share an entangled pair of qubits that could then be used for quantum communications. In this work we show that this ES scheme is tolerant to low levels of loss in the propagating coherent states, resulting in Alice and Bob ultimately sharing a pair of highly entangled qubits of impressive fidelity when compared to the maximally entangled $\ket{\Phi^{+}}$ Bell state.
	
	In this ES scheme we have a beam-splitter (BS) of transmission $T$ described by $BS^{T}_{i,j}$, where $i$ and $j$ are the modes that are mixed at the BS. Let us therefore assume that we have a loss rate of $1-T$ in a channel, modelled by mixing modes $B$ and $D$ with vacuum states in modes $\varepsilon_{B}$ and $\varepsilon_{D}$ respectively at separate BSs. Each hybrid entangled state is then given by 
		\begin{align}\label{initialstate}
	\ket{\psi_{loss}}_{AB\varepsilon_{B}}=BS^{T}_{B,\varepsilon_{B}}\ket{\psi_{HE}}_{AB}\ket{0}_{\varepsilon_{B}}\nonumber\\
	=\frac{1}{\sqrt{2}}\Big(\ket{0}_{A}\ket{\alpha\sqrt{T}}_{B}\ket{\alpha\sqrt{1-T}}_{\varepsilon_{B}}+\nonumber\\
	\ket{1}_{A}\ket{-\alpha\sqrt{T}}_{B}\ket{-\alpha\sqrt{1-T}}_{\varepsilon_{B}}\Big),
	\end{align}
	\noindent where the hybrid entangled quantum state is given in Eq. \ref{hybrid}. Note that Eq. \ref{initialstate} is identical for modelling loss in mode $D$, using a vacuum state in mode $\varepsilon_{D}$.
	
	After accounting for loss as described above, we then mix the two propagating lossy modes at a 50:50 BS. Mixing two coherent states with a (generalised) BS of transmission $t$ is given by 
	\begin{flalign}
	BS^{t}_{B,D}&\ket{\alpha}_{B}\ket{\beta}_{D}=\nonumber\\
	&\ket{\alpha\sqrt{t}-\beta\sqrt{1-t}}_{B}\ket{\alpha\sqrt{1-t}+\beta\sqrt{t}}_{D},&
	\end{flalign} 
	\noindent where, $\alpha$ and $\beta$ are complex numbers. In this protocol we mix coherent states of the same amplitude using a 50:50 BS, therefore $t=1/2$.	
	
	\subsection{Detection Methods}\label{Detection}
	For successful ES, we measure mode $D$ via (perfect) balanced homodyne detection, and mode $B$ by a vacuum measurement. It was found that if two homodyne measurements are performed on modes $B$ and $D$, then the resultant quantum state is a superposition of all possible 2 qubit strings, which is a product state and is therefore undesirable as an outcome for this protocol. 
	
\begin{figure}[H]
		\centering
		\scalebox{0.5}{\includegraphics[trim={2cm 2cm 2cm 2.5cm},clip]{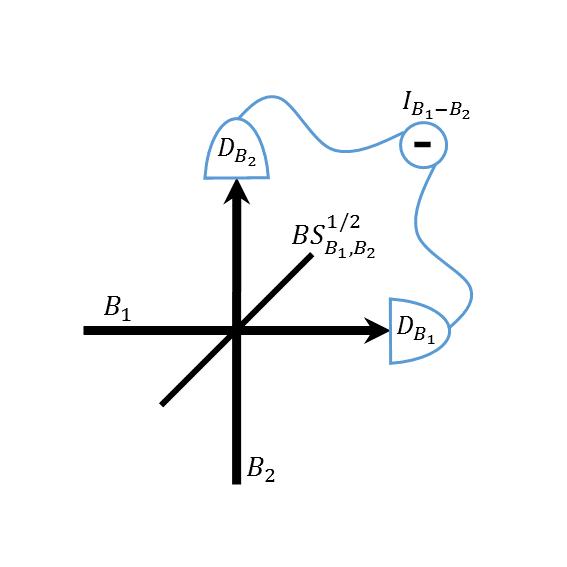}}
		\caption{Diagram to represent the two channel system undergoing balanced homodyne detection, where $B_1$ is the input signal (mode $D$ in protocol), and $B_2$ is the local oscillator. $I_{B_{1}-B_{2}}$ is the intensity difference between the photodetectors $D_{B_{1}}$ and $D_{B_{2}}$.}
		\label{fig:Homodyne}
	\end{figure}

	A generalised scheme of balanced homodyne detection consists of one 50:50 BS, a strong coherent field $\ket{\beta e^{i\theta}}$ of amplitude $\beta$ (where $\beta$ is real) and two photodetectors; the probe mode (mode $D$) is combined at a BS with the strong coherent field (\enquote{local oscillator}) of equal frequency, and photodetection is then used to measure the outputs \cite{Scully1997}. If we perform homodyne detection on an input signal in mode $B_{1}$ and the coherent field is injected in mode $B_{2}$, then the operator $BS^{1/2}_{B_{1},B_{2}}$ mixes the input state and the coherent field, as shown in Fig. \ref{fig:Homodyne}.
	The intensity difference (photon number difference) between the two photodetectors ($D_{{B}_{1}}$ and $D_{{B}_{2}}$) can be calculated using the two mode operator $\hat{I}_{B_{1}-B_{2}}=\hat{b}^{\dagger}_{1}\hat{b}_{2}+\hat{b}^{\dagger}_{2}\hat{b}_{1}$, with creation and annihilation operator denoted by $\hat{b}^{\dagger}_{i}$ and $\hat{b}_{i}$ respectively, in mode $B_{i}$. It therefore follows that, 
	\begin{align}
	\hat{I}_{B_{1}-B_{2}}=2\beta\langle\hat{x}_{\theta}\rangle,
	\end{align}
	\noindent where, $\hat{x}_{\theta}=\frac{1}{2}\left(\hat{b}_{1}e^{-i\theta}+\hat{b}^{\dagger}_{1}e^{i\theta}\right)$ \cite{Gerry2005}, $\beta$ is the amplitude of the strong coherent field injected in mode $B_{2}$, and the phase of the quadrature $\hat{x}_{\theta}$ is given by the phase of this local oscillator. The probability amplitude of a homodyne measurement on an arbitrary coherent state $\ket{\alpha e^{i\varphi}}$ can be described by projecting with an $\hat{x}_{\theta}$ eigenstate, where $\hat{x}_{\theta}\ket{x_{\theta}}=x_{\theta}\ket{x_{\theta}}$, for real $\alpha$ \cite{Gardiner2004}:
	
	\begin{align}
	\braket{{x_{\theta}}|{\alpha e^{i\varphi}}}=\frac{1}{\pi^{\frac{1}{4}}}\exp\Bigg[-\frac{1}{2}(x_{\theta})^{2}+\sqrt{2}e^{i(\varphi-\theta)}\alpha x_{\theta}\nonumber\\
	-\frac{1}{2}e^{2i(\varphi-\theta)}\alpha^{2}-\frac{1}{2}\alpha^{2}\Bigg],
	\end{align}
	\noindent where the subscript on $x_{\theta}$ is indicative of the angle in which the homodyne measurement is performed. In this protocol specifically we will theoretically measure mode $D$ using homodyne detection in the $\theta=\frac{\pi}{2}$ plane; if we measure in this plane then we are not able to distinguish between the two remaining states ($\ket{00}_{AC}$ and $\ket{11}_{AC}$), thus leaving them entangled, whereas if one were to measure in the $\theta=0$ plane then these states are distinguishable, which would destroy any entanglement. 
	
	\subsection{Entanglement Swapping with Equal Lossy Modes}\label{Equal}
	Measuring a vacuum in mode $B$ and performing homodyne detection in mode $D$ results in the following state, which shows the entangled pair of qubits shared by Alice and Bob after carrying out this protocol in its entirety (prior to tracing out the lossy modes):	
	\begin{align}\label{finalstate}
	&\ket{\psi_{loss}}_{AC\varepsilon_{B}\varepsilon_{D}}=\nonumber\\
	&Ne^{(T-1)|\alpha|^{2}}\sum_{n,m=0}^{\infty}\frac{(\alpha\sqrt{1-T})^{n+m}}{\sqrt{n}!\sqrt{m}!}\ket{n}_{\varepsilon_{B}}\ket{m}_{\varepsilon_{D}}.\nonumber\\
	&\Big(e^{-2i\alpha x_{\frac{\pi}{2}}\sqrt{T}}\ket{00}_{AC}+(-1)^{n+m}e^{2i\alpha x_{\frac{\pi}{2}}\sqrt{T}}\ket{11}_{AC}\nonumber\\
	&+e^{-T|\alpha|^{2}}((-1)^{m}\ket{01}_{AC}+(-1)^{n}\ket{10}_{AC})\Big),
	\end{align}
	\noindent where, $N$ is a normalisation coefficient, and the lossy modes ($\varepsilon_{B}$ and $\varepsilon_{D}$) are summed over $\ket{n}_{\varepsilon_{B}}$ and $\ket{m}_{\varepsilon_{D}}$ respectively (using the Fock (number) state basis representation of a coherent state, $\ket{\alpha}=e^{-\frac{|\alpha|^{2}}{2}}\sum_{n=0}^{\infty}\frac{\alpha^{n}}{\sqrt{n}!}\ket{n}$ \cite{Gerry2005}). If one sets the amplitude of the coherent state as $T|\alpha|^{2}>>1$ in Eq. \ref{finalstate}, then the resultant state contains only the diagonal $\ket{00}_{AC}$ and $\ket{11}_{AC}$ terms, as the off-diagonal $\ket{01}_{AC}$ and $\ket{10}_{AC}$ terms are rapidly exponentially dampened by the exponent $e^{-T|\alpha|^{2}}$. After tracing out the lossy modes, and taking these limits of $T|\alpha|^{2}>>1$, the resultant density matrix from this quantum state is
	\begin{align}\label{optimumstate}
	\rho_{AC}\approx&\frac{1}{2}e^{2(T-1)|\alpha|^{2}}\sum_{n,m=0}^{\infty}\frac{((T-1)\alpha^{2})^{n+m}}{n!m!}.\nonumber\\
	&\Big[\ket{00}_{AC}\bra{00}+\ket{11}_{AC}\bra{11}\nonumber\\
	&+(-1)^{n+m}e^{4i\alpha x_{\frac{\pi}{2}}\sqrt{T}}\ket{11}_{AC}\bra{00}\nonumber\\
	&+(-1)^{n+m}e^{-4i\alpha x_{\frac{\pi}{2}}\sqrt{T}}\ket{00}_{AC}\bra{11}\Big].
	\end{align}  
	Note that the phase factors in Eq. \ref{optimumstate} are known phase factors, set by the measurement outcome $x_{\frac{\pi}{2}}$. These can either be corrected through local operations feeding forward the measurement result, or simply carried through the protocol and dealt with in subsequent post-processing. 
	
	It will be shown in Section \ref{RandD} that the entanglement negativity, fidelity and linear entropy of $\rho_{AC}$, with respect to the maximally entangled Bell State $\ket{\Phi^{+}}=\frac{1}{\sqrt{2}}(\ket{00}+\ket{11})$, is optimal for a specific value of the amplitude ($\alpha$) of the coherent states that propagate through the lossy modes. 
	
	\section{Entanglement Swapping with Unequal Lossy Modes}\label{Unequal}
	
	It is important to consider the case of unequal lossy modes in this protocol; in reality the beam-splitters used to mimic lossy optical fibres will not be absolutely equal, the resultant states that are emitted will have different transmission ($T$) values. However, we show here that the entanglement shared between Alice and Bob after performing ES is not significantly damaged if we consider unequal loss.
	
	Firstly, we denote this \enquote{loss mismatch} variable as $\delta$, and we parametrise the transmission in each lossy mode as ${T_{B}}\rightarrow{T}$ and ${T_{D}}\rightarrow{T-\delta}$ where, like $T$, $\delta$ can only take a value between 0 and 1. In general $\delta$ will be a small, positive mismatch to avoid $T_{D}$ exceeding unity. Performing an analogous derivation to that used to reach Eq. \ref{finalstate}, and applying the above parametrisation gives  
	\begin{align}\label{finalfinalstate}
	\ket{\psi^{loss}}_{AC\varepsilon_{B}\varepsilon_{D}}=Ne^{(T-\frac{\delta}{2}-1)|\alpha|^{2}}.\nonumber\\
	\sum_{n,m=0}^{\infty}\frac{(\alpha\sqrt{1-T})^n(\alpha\sqrt{1-T+\delta})^m}{\sqrt{n}!\sqrt{m}!}&\ket{n}_{\varepsilon_{B}}\ket{m}_{\varepsilon_{D}}.\nonumber\\
	\Big(e^\frac{-|\alpha\mathcal{T}_{-}|^{2}}{4}e^{-\mathcal{T}_{+}i\alpha x_{\frac{\pi}{2}}}&\ket{00}_{AC}\nonumber\\
	+(-1)^{m}e^\frac{-|\alpha\mathcal{T}_{+}|^{2}}{4}e^{-\mathcal{T}_{-}i\alpha x_{\frac{\pi}{2}}}&\ket{01}_{AC}\nonumber\\
	+(-1)^{n}e^\frac{-|\alpha\mathcal{T}_{+}|^{2}}{4}e^{\mathcal{T}_{-}i\alpha x_{\frac{\pi}{2}}}&\ket{10}_{AC}\nonumber\\
	+(-1)^{n+m}e^\frac{-|\alpha\mathcal{T}_{-}|^{2}}{4}e^{\mathcal{T}_{+}i\alpha x_{\frac{\pi}{2}}}&\ket{11}_{AC}\Big),
	\end{align}
	where, ${\cal T}_{\pm}=(\sqrt{T}\pm \sqrt{T-\delta})$.  As an example here we consider the case where a system is set up for matched loss $(1-T)$ but there is a small, unknown mismatch. This can be calculated by taking an average over a distribution of $\delta$.
	To find the averaged density matrix ($\overline{\rho}_{AC}$) of the state $\ket{\psi^{loss}}_{AC\varepsilon_{B}\varepsilon_{D}}$, for some width in the distribution of the loss mismatch $\delta$, which we label as $\Delta$, we must integrate the density matrix $\rho_{AC}(\delta,T,\alpha)$ over all positive values of $\delta$ (where $\rho_{AC}(\delta,T,\alpha)=\ket{\psi^{loss}}_{AC\varepsilon_{B}\varepsilon_{D}}\bra{\psi^{loss}}$). The distribution of the loss mismatch is a one-sided (positive) Gaussian curve, and so the integral is of the form 
	\begin{eqnarray}\label{AveragedRho}
	\overline{\rho}_{AC}\equiv\int_{0}^{\infty}f(\delta,\Delta)\rho_{AC}(\delta,T,\alpha)d\delta,
	\end{eqnarray}
	\noindent where, $f(\delta,\Delta)=\sqrt{\frac{2}{\pi\Delta^{2}}}e^{\frac{-\delta^{2}}{2\Delta^{2}}}$ and $\sqrt{\frac{2}{\pi\Delta^{2}}}$ is the normalisation of the function. We will show in the next section that this averaged density matrix provides a high level of entanglement for an optimum $\alpha$ value when considering low levels of loss, and unequal loss in modes $\varepsilon_{B}$ and $\varepsilon_{D}$.
	
	We note that equation (10) could be used directly to model a known mismatch between losses (for example due to unequal lengths of fibre), by choosing a specific value of $\delta$. The results of such calculations show very similar impact on the entanglement to those we give for averaging with a width $\Delta$, so we do not present these.
	
	\section{Results and Discussion}\label{RandD}
	
	The fidelity ($F$) of the final density matrix (Eq. \ref{AveragedRho}) can be determined using 
	\begin{eqnarray}
	F\left(\ket{\sigma},\rho\right)=\bra{\sigma}\rho\ket{\sigma},
	\end{eqnarray} 
	\noindent where, $\ket{\sigma}=\ket{\Phi^{+}}$ is the maximally entangled (pure) Bell State, and $\rho=\overline{\rho}_{AC}$ is the final averaged density matrix \cite{Jozsa1994}. Calculating the closeness (fidelity) of $\overline{\rho}_{AC}$ to $\ket{\Phi^{+}}$ confirms that for an optimum amplitude of the coherent state ($\alpha\approx1.5$), $T=0.99$ in mode $B$ and $T=0.98$ in mode $D$, the final state shared by Alice and Bob is of impressive fidelity: $F=0.93$, where a fidelity of $F=1$ indicates that the states in comparison are indistinguishable. Intrinsically, the fidelity is unity for the no loss case, but what is promising here is that even for the case with non-negligible loss where $T=0.95$ in mode $B$ and $T=0.94$ in $D$ the fidelity reaches a maximum of 0.81 for $\alpha=1.3$. 
		
	To evaluate the level of entanglement shared between Alice and Bob after performing entanglement swapping, we apply an entanglement measure called \enquote{negativity} \cite{Vidal2002} using the following:
	\begin{eqnarray}
	E\left(\overline{\rho}_{AC}\right)=-2\sum_{i}^{}\lambda_{i}^{-},
	\end{eqnarray}
	\noindent where $E$ denotes the entanglement value of $\overline{\rho}_{AC}$ (which can take a value between 0, for no entanglement, and 1, for maximal entanglement), and $\lambda_{i}^{-}$ represents the negative eigenvalues of the partial transpose of the final density matrix, $\overline{\rho}_{AC}$. We also calculate the linear entropy of $\overline{\rho}_{AC}$ using
	\begin{eqnarray}
	S_{L}\left(\overline{\rho}_{AC}\right)=1-\text{Tr}\left[{\overline{\rho}_{AC}}^{2}\right],
	\end{eqnarray}
	\noindent where $S_{L}$ is the linear entropy of the system, and can take any value between 0 (for a pure state) to $S_{L}^{max.}=1-\frac{1}{d}$, where $d$ is the dimension of the system \cite{Manfredi2000}. Therefore, in this case the maximum linear entropy will be 0.5, corresponding to a maximally mixed state. 
	
	The following plots show entanglement and linear entropy as a function of the amplitude ($\alpha$) of the coherent states used, with fixed transmission ($T$) values.
	
	\end{multicols}
	\begin{figure}[H]
	\subfloat[]{\scalebox{0.3}{\includegraphics{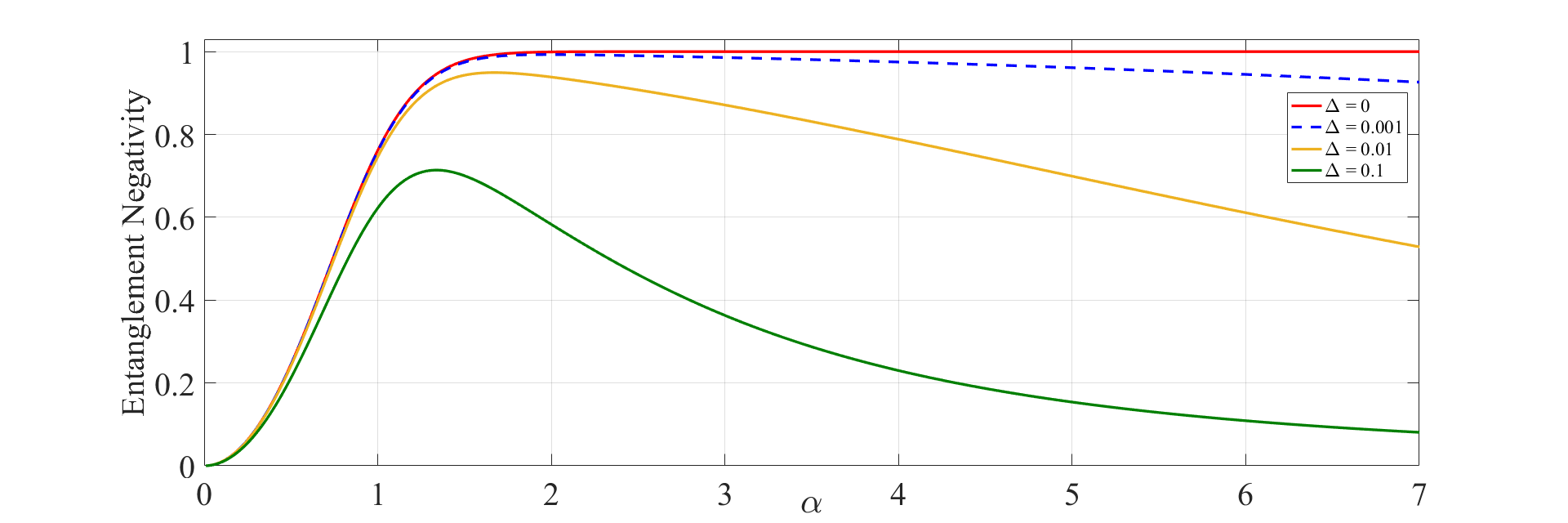}}
	\label{fig:EntanglementT100}}%
	
	\subfloat{
	\scalebox{0.3}{\includegraphics{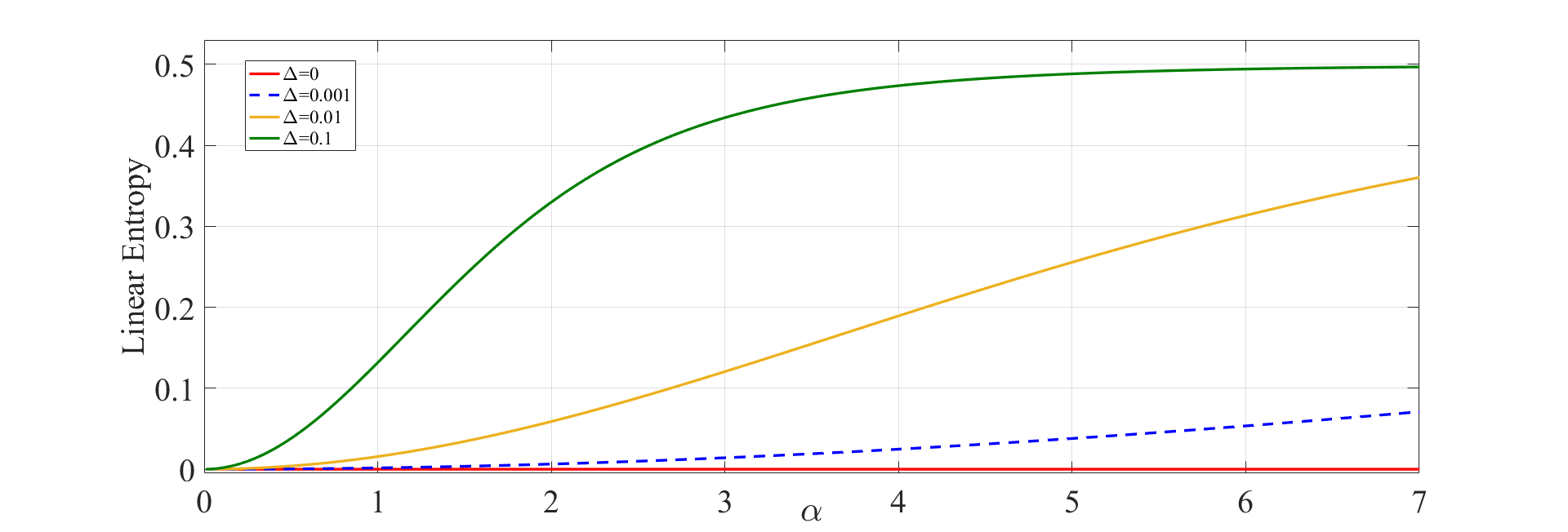}}
	\label{fig:EntropyT100}}%
	
	\caption{Plot of entanglement (\ref{fig:EntanglementT100}) and linear entropy (\ref{fig:EntropyT100}) as a function of $\alpha$, for a transmission of $T=1$ and $\Delta=0,0.001,0.01,0.1$.}
	\end{figure}

	\begin{figure}[H]
	\subfloat[]{\scalebox{0.3}{\includegraphics{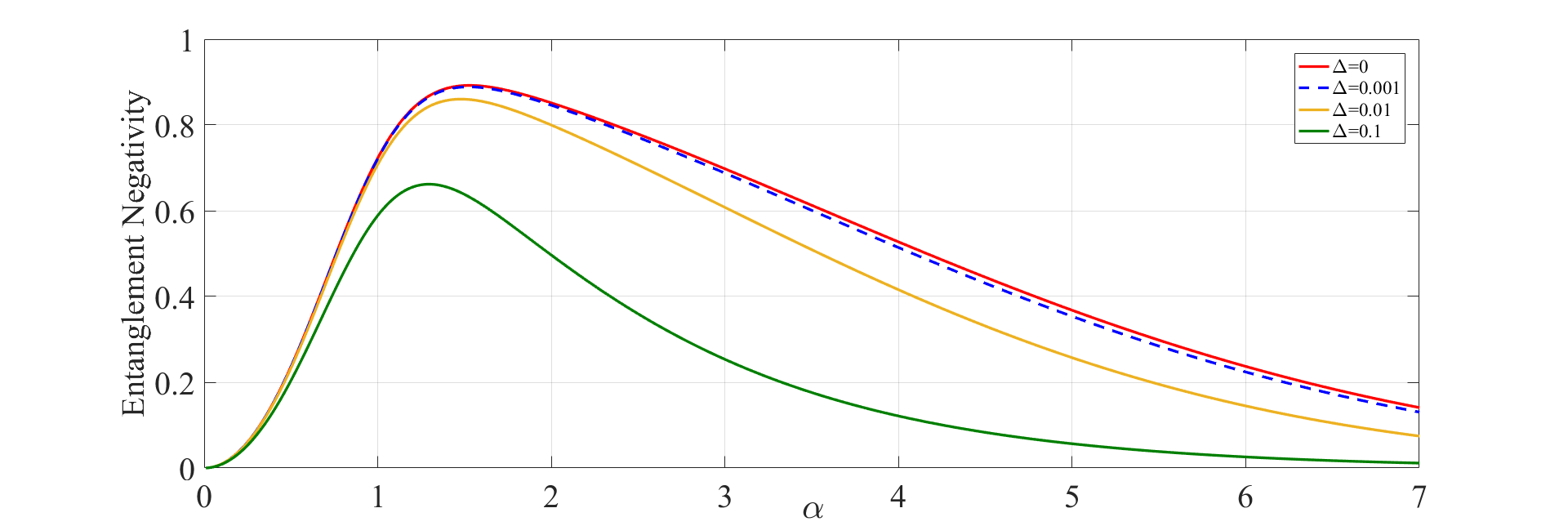}}
	\label{fig:EntanglementT99}}%

	\subfloat[]{\scalebox{0.3}{\includegraphics{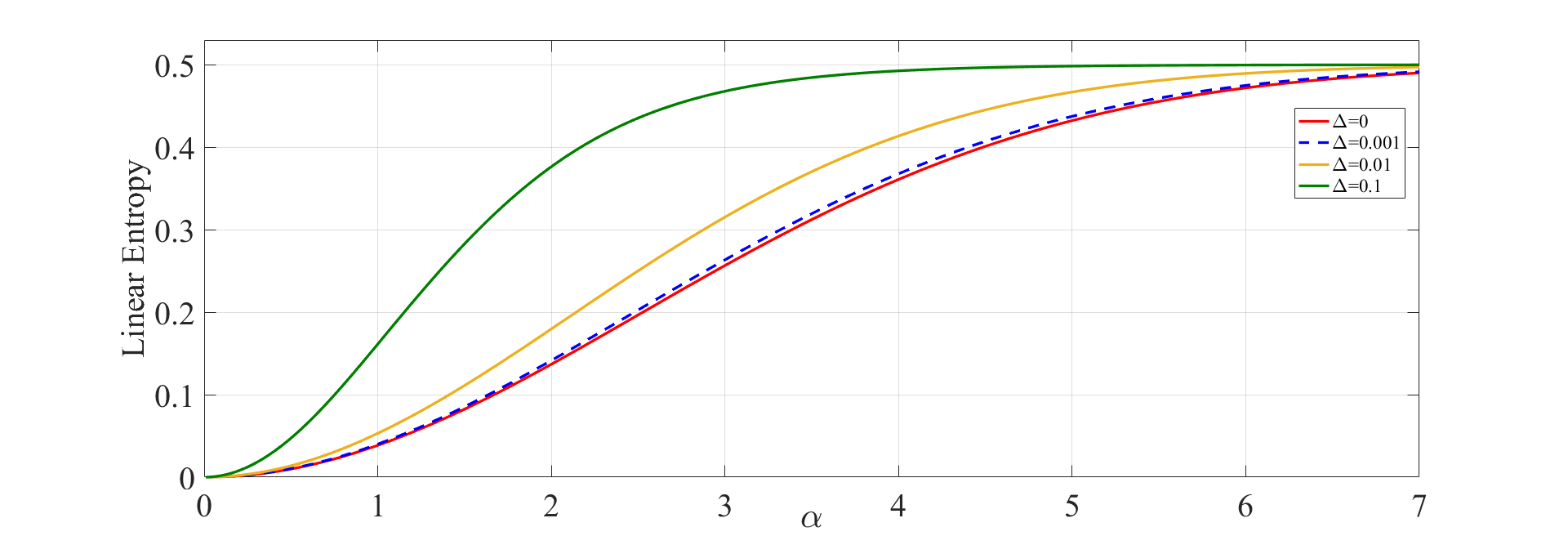}}	
	\label{fig:EntropyT99}}%

	\caption{Plot of entanglement (\ref{fig:EntanglementT99}) and linear entropy (\ref{fig:EntropyT99}) as a function of $\alpha$, for a transmission of $T=0.99$ and $\Delta=0,0.001,0.01,0.1$.}
	\end{figure}

	\begin{figure}[H]
	\subfloat[]{\scalebox{0.3}{\includegraphics{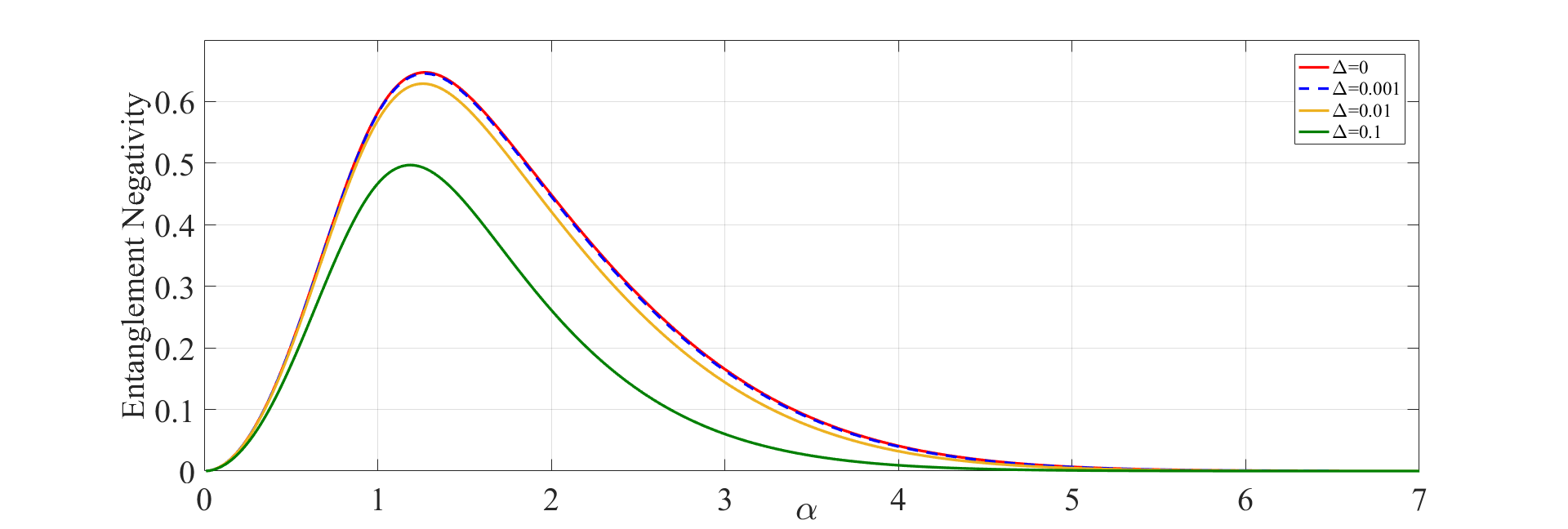}}
	\label{fig:EntanglementT95}}%

	\subfloat[]{\scalebox{0.3}{\includegraphics{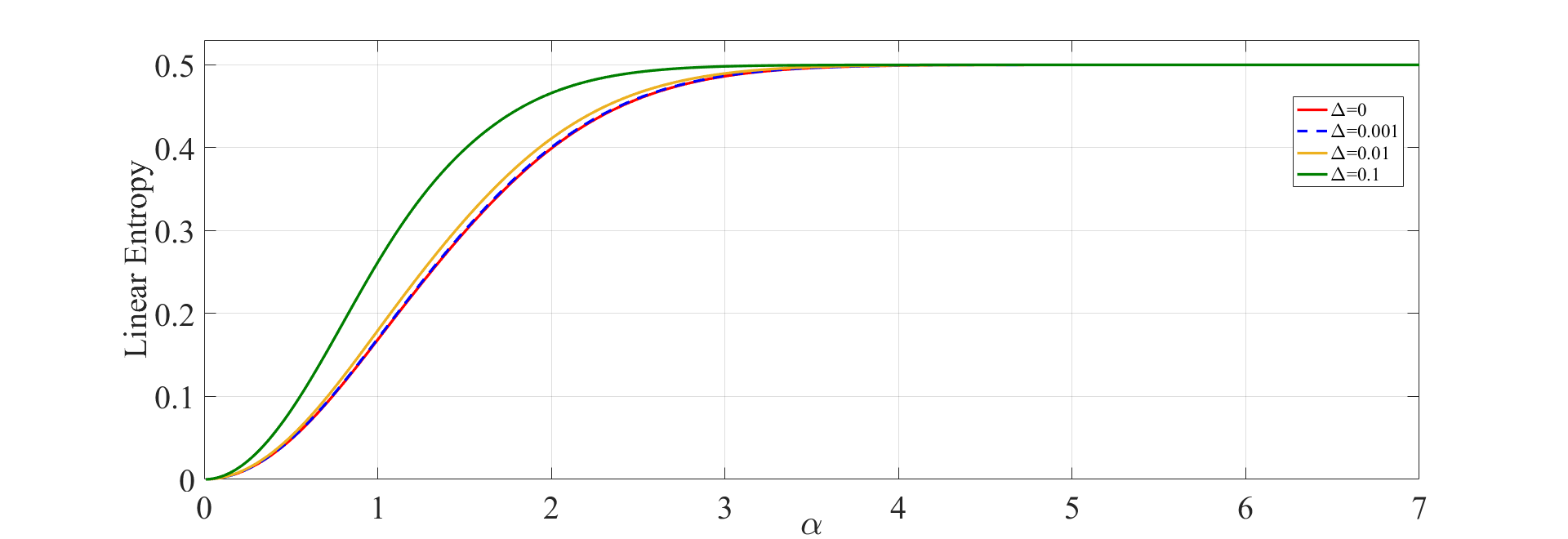}}	
	\label{fig:EntropyT95}}%

	\caption{Plot of entanglement (\ref{fig:EntanglementT95}) and linear entropy (\ref{fig:EntropyT95}) as a function of $\alpha$, for a transmission of $T=0.95$ and $\Delta=0,0.001,0.01,0.1$.}
	\end{figure}
 \begin{multicols}{2}

Fig. \ref{fig:EntanglementT100} shows that for no loss in the system ($T=1,\Delta=0$), the entanglement reaches unity when $\alpha>1.7$. For finite loss, when $T<1$, the optimum value of entanglement is approached and is clearly given by a sharp peak as a function of $\alpha$ (see Figs. \ref{fig:EntanglementT99} and \ref{fig:EntanglementT95}). Although this shifts to slightly lower values of $\alpha$ when considering higher levels of loss, there is always a clear peak in the plot at a specific amplitude. This is as a result of the analytical expression defining the shared state between Alice and Bob (Eq. \ref{finalfinalstate}), where the off-diagonal states (with the exception of $\ket{00}_{AC}\bra{11}$ and $\ket{11}_{AC}\bra{00}$) are dampened when $\alpha>>0$. This therefore reduces the entanglement, and also explains why the plots tail off at higher amplitudes for finite $T$.

This is a key point of this paper: to have an optimum $\alpha$ value means that for a practical demonstration of this protocol an experimentalist would know the level of loss that can be tolerated, given the amplitude of the coherent state they have prepared. Furthermore, this optimum value itself is desirable - an amplitude of 2 is not large, but importantly it also is not too close to a vacuum state as to be indistinguishable. Equally, were the amplitudes of the coherent states to be closer to 0 then there is the possibility that these states will overlap at the vacuum, therefore making the superposition of $\ket{\alpha}_{D}$ and $\ket{-\alpha}_{D}$ indistinguishable in a homodyne measurement. Again, this further proves the possibility of performing this protocol experimentally, as a coherent state of this kind of amplitude can be prepared experimentally.  

When $T=0.95$, Fig. \ref{fig:EntanglementT95} shows that even when considering high levels of loss for unequal lossy modes ($\Delta=0.01$) the entanglement value is 0.63 for $\alpha\approx1.3$. Although the  state shared by Alice and Bob is not highly entangled in this case it is nonetheless still useful as a proof-of-principle experiment of this particular entanglement swapping protocol. What is promising in this protocol is that in Fig. \ref{fig:EntanglementT99}, for a transmission of $T=0.99$ in one mode and $T=0.98$ in the other ($\Delta=0.01$) the maximum entanglement value is 0.87, for $\alpha\approx1.5$; these levels of loss are likely to be the most realistic case for a practical implementation of this protocol, and although the entanglement is slightly lessened as a result of this loss, there do still exist methods of increasing entanglement, such as entanglement purification schemes \cite{Bennett1996,Christoph2001,Sheng2014}. 

The linear entropy plots compliment the plots of entanglement as a function of $\alpha$ perfectly: it is clear from comparing linear entropy and entanglement plots of the same transmission value that as entanglement increases as function of $\alpha$, the linear entropy decreases for the same amplitude. What is also worth noting is that in all linear entropy plots, the case where we have significant differences in the lossy modes ($\Delta=0.1$) gives the plots that show the highest level of entropy in the system. This of course arises from the unequal lossy modes causing the overall quantum state shared by Alice and Bob to be more mixed, which in turn is confirmed by the entanglement plots showing lower levels of entanglement for $\Delta=0.1$. 

Another important quantity to evaluate is the success probability of the protocol. Here we focus on the success probability of the vacuum projection (in mode $B$) in this ES scheme. Clearly what is of interest is the success probability where the entanglement peaks as a function of the coherent state amplitude $\alpha$. Calculation of this success probability shows that it is unity for the case of very small $\alpha$, but drops rapidly and plateaus at 1/2 at the same value of $\alpha$ where the entanglement plots peak ($\alpha\approx1.5$). What is promising here is that the success probability does not decrease as $T$ drops from 1 to 0.95. Furthermore, the loss mismatch does not reduce the success probability in the regime of small $\alpha$, and only drops to less than 1/2 when $\alpha>3$, for a significant mismatch in loss ($\Delta=0.1$). Note that as we are assuming a perfect homodyne detection scheme the success probability will inherently be unity in this case. Investigating imperfect homodyne detection will be interesting as future work. 

\section{Conclusions}\label{Conclusions}
	Crucial to this scheme is that the measurements outlined in Section \ref{Detection} must be performed specifically as stated (that is, a vacuum projection in mode $B$ and a homodyne detection in $D$). In doing so, one can theoretically achieve high levels of entanglement for low levels of photon loss. There are three key points to this paper which are worth summarising once more: \\[0.1in]
	
	\begin{itemize}
		\item Having unequal loss does not significantly impact the entanglement and fidelity values, and the protocol is actually fairly resilient to this
		\item We can reach optimum entanglement, fidelity and linear entropy for a specific value of the amplitude ($\alpha$) of the propagating coherent states 
		\item The most realistic (practical) case is a transmission of $T=0.99$, and a loss mismatch of $\Delta=0.01$, resulting in an impressive entanglement value of 0.87 for $\alpha\approx1.5$
	\end{itemize}

	\noindent This work highlights the usefulness of entangled optical hybrid states of light, and shows that the continuous variable part of this hybrid state is particularly resilient to low levels of photon losses. Furthermore, if applied with a suitable entanglement purification scheme \cite{Sheng2013}, this protocol has the potential to be implemented as part of a full quantum repeater protocol. Under the assumption of small losses in a channel, the ES protocol could also be used for entangling two distant superconducting qubits. These can be entangled because the state $\ket{\psi_{HE}}$ can easily be created between a superconducting qubit and a coherent state inside a superconducting circuit \cite{Xiang2017}. 
	 
	Further work includes investigating cat states (coherent state superpositions) as the propagating continuous variable in the hybrid state, and also investigating the impact of imperfect homodyne detection to this entanglement swapping protocol. 
 
	\section{Acknowledgements} 
	We acknowledge support from EPSRC (EP/M013472/1). J.J. acknowledges support from the KIST Institutional Program (Project No. 2E26680-16-P025). 
	 
	 \end{multicols}

\end{document}